\documentstyle[12pt,aaspp4,psfig]{article}

\begin{document}

\title{Two Rare Magnetic Cataclysmic Variables with Extreme Cyclotron
Features Identified in the Sloan Digital Sky
Survey\altaffilmark{1}}

\author{Paula Szkody\altaffilmark{2}, Scott F. Anderson\altaffilmark{2},
Gary Schmidt\altaffilmark{3}, 
Patrick B. Hall\altaffilmark{4},
Bruce Margon\altaffilmark{5},
Antonino Miceli\altaffilmark{2},
Mark SubbaRao\altaffilmark{6},
James Frith\altaffilmark{2},
Hugh Harris\altaffilmark{7},
Suzanne Hawley\altaffilmark{2}, Brandon Lawton\altaffilmark{2},
Ricardo Covarrubias\altaffilmark{2},
Kevin Covey\altaffilmark{2},  Xiaohui Fan\altaffilmark{8}, 
Thomas Murphy\altaffilmark{2}
Vijay Narayanan\altaffilmark{4},
Sean Raymond\altaffilmark{2}, 
Armin Rest\altaffilmark{2},
Michael A. Strauss\altaffilmark{4},
Christopher Stubbs\altaffilmark{2}, Edwin Turner\altaffilmark{4},
Wolfgang Voges\altaffilmark{9}, 
Amanda Bauer\altaffilmark{6},
J. Brinkmann\altaffilmark{10}, Gillian R. Knapp\altaffilmark{4},
Donald P. Schneider\altaffilmark{11} }

\altaffiltext{1}{Based in part on observations obtained with the Sloan Digital
Sky Survey and with the Apache Point Observatory (APO) 3.5m telescope, which
are owned and operated by the Astrophysical Research Consortium (ARC).  A
portion of the observations reported here were obtained at the MMT
Observatory, a joint facility of the University of Arizona and the Smithsonian
Institution.}
\altaffiltext{2}{Department of Astronomy, University of Washington, Box 351580,
Seattle, WA 98195}
\altaffiltext{3}{The University of Arizona, Steward Observatory, Tucson, AZ 85721}
\altaffiltext{4}{Princeton University Observatory, Peyton Hall, Princeton, NJ 08
544}
\altaffiltext{5}{Space Telescope Science Institute, 3700 San Martin Drive, Baltimore, MD 21218}
\altaffiltext{6}{Astronomy and Astrophysics Center, University of
Chicago, 5640 South Ellis Avenue, Chicago, IL 60637}
\altaffiltext{7}{US Naval Observatory, Flagstaff, AZ 86002}
\altaffiltext{8}{The Institute for Advanced Study, Princeton, NJ 08540}
\altaffiltext{9}{Max-Planck-Institute f\"ur extraterrestrische Physik,
Geissenbachstr. 1, D-85741 Garching, Germany}
\altaffiltext{10}{Apache Point Observatory, PO Box 59, Sunspot, NM 88349}
\altaffiltext{11}{Department of Astronomy \& Astrophysics, Penn State University, University Park, PA 16802}

\begin{abstract}
Two newly identified magnetic cataclysmic variables discovered in the Sloan
Digital Sky Survey (SDSS), SDSSJ155331.12+551614.5 and SDSSJ132411.57+032050.5,
have spectra showing highly prominent, narrow, strongly polarized
 cyclotron humps with 
amplitudes that vary on orbital periods of 4.39 and 2.6 hrs, respectively. 
In the former, the spacing
of the humps indicates the 3rd and 4th harmonics in
 a magnetic field of $\sim$60MG. The narrowness of the cyclotron features
and the lack of strong emission lines imply very low temperature plasmas and
very low accretion rates, so that the accreting area is heated by particle
collisions rather than accretion shocks.
The detection of rare systems like these 
exemplifies the ability of the SDSS
to find the lowest accretion rate close binaries.
\end{abstract}


\section{Introduction}

The commissioning year of the Sloan Digital Sky Survey (SDSS; York et al. 2000,
Stoughton et al. 2002)
 showed that
this survey is highly effective in finding new cataclysmic variables (Szkody et al. 2002).
While previous surveys had primarily identified the brightest systems with the
highest mass transfer rates, the SDSS photometry in 5 filters
to well beyond 20th magnitude (Gunn et al. 1998, Fukugita et al. 1996, Hogg et
al. 2001, Pier et al. 2002, Smith et al. 2002) is able
to find the population of low mass transfer rate, very short orbital period
systems that are predicted to exist in close binary evolution models (Howell,
Rappaport \& Politano 1997). Included in this latter group are some systems
which contain highly magnetic white dwarfs with field strengths 6-240 MG which
are termed AM Her systems, or polars (see Warner 1995 for a review
of all types of cataclysmic
variables).

In the polars, the high field prevents the formation of an accretion disk by
directing the ballistic flow of
material transferred from
the late type secondary onto one or both magnetic poles of the white dwarf.
Different regimes of accretion rate
and magnetic field strength are predicted to result in quantitatively
different accretion scenarios  
 (Wickramasinghe \& Ferrario
2000; WF2000).
At the highest specific accretion rates (100 g cm$^{-2}$ s$^{-1}$),
high density blobs carry the mass and energy
below the surface of the white dwarf. At mid accretion rates
(1 g cm$^{-2}$ s$^{-1}$), a standoff
shock is formed above the surface and the gas cools primarily by 10-30 keV
 thermal
bremsstrahlung emission. As the accretion lowers by another factor of 100,
the cooling becomes dominated by cyclotron emission, until at the lowest
rates, a shock does not form at all and the energy of the incoming ions is
transmitted directly to the atmosphere of the white dwarf in what is termed
a ``bombardment solution" (Kuipers \& Pringle 1982; Fischer \& Beuermann 2001).
 The magnetic field 
also affects
the results in that higher fields tend to produce weaker shocks and possibly 
more
direct heating by blobs beneath the surface (Ramsay et al. 1994). Complicating
this picture further is the fact that the accretion rate of polars can 
sporadically
change. Typically, sytems in high mass-transfer states exhibit 
strong HeII $\lambda$4686 and Balmer emission lines
which dominate the optical spectrum, and strong X-ray and cyclotron
emission. At low states of reduced or absent mass
transfer, the line emission disappears (except for some narrow Balmer emission
from the irradiated secondary star), the X-ray and cyclotron emission is much
reduced or absent, and the photospheres of the white dwarf and late secondary
are visible.

During low states, the magnetic field may be identified through Zeeman
splitting in the photospheric spectrum, but for most systems, the field is
determined 
in high states of
accretion from the presence
of cyclotron harmonics in the optical and near-IR. Cyclotron opacity is a
rapidly decreasing function of harmonic number, and the high harmonics are
strongly angle-dependent, polarized, and broadened with increasing electron 
 temperature (e.g. Chanmugam 1980; Meggitt \& Wickramasinghe 1982). For
T$<<$10 keV, the wavelength of the harmonic number {\it n} is simply related
 to the magnetic field $B$:
$\lambda_n = (10,700)/n)(10^{8}/B)$ \AA.

Within the above theoretical and empirical framework, 
observation of the soft and hard X-ray
fluxes, and of the optical spectral and cyclotron features, can elucidate the
magnetic field and accretion regime of identified polars. More than 80\%
of the $\sim$ 65
known polars were discovered in the ROSAT All Sky Survey (RASS; Voges
et al. 1999),
with typical count rates for 15-19th optical magnitudes in the range of
0.2-2.5 c/s (Beuermann \& Burwitz 1995). The selection criteria of
high X-ray count rate and
strong optical emission lines resulted in the discovery of polars in 
the middle to high specific accretion rate
regimes. Recently, the deep objective prism plates of the Hamburg Quasar survey
provided the identification of 2 polars with extremely low accretion rates 
(WX LMi = HS 1023+3900;
Reimers, Hagen \& Hopp 1999, and HS 0922+1333; Reimers \& Hagen 2000) and 
the followup of faint ROSAT sources yielded two more (RX J012851.9-233931; 
Schwope, Schwarz \& Greiner 1999, and RX J1007.5-2016; Reinsch et al. 1999).
 In this paper, we describe 2 new SDSS polars which are among the most extreme
cases of the intriguing
 cyclotron-dominated systems at the lowest accretion rates.
As only a small fraction of the eventual SDSS data have been examined, the
survey should discover a modest-sized sample of these previously exotic stars.

\section{Observations}

The objects SDSSJ155331.12+551614.5 and
SDSSJ132411.57+032050.5 (hereafter abbreviated as SDSS1553 and SDSS1324)
were automatically selected for SDSS spectroscopy
since their unusual SDSS colors ($r$=17.43, $u-g$=1.52, $g-r$=1.06, $r-i$=0.40, 
$i-z$=1.00 for SDSS1553 and $r$=20.44, $g-r$=1.65, $r-i$=0.21, $i-z$=0.83,
$u-g\sim$1.1 for SDSS1324
with no reddening correction) 
fell far from the stellar locus and within
the selection criteria for quasars (Richards et al. 2002). 
The SDSS spectra with $\sim$3\AA\ resolution (Figure 1),
show strong broad cyclotron features at 4600\AA\ and 6200\AA\ in SDSS1553 and
at 5700\AA\ in SDSS1324 together with the TiO band features of late-type 
main-sequence stars.
Periodic photometric
variability at different wavelengths (Figure 2) was used to determine the
orbital periods. Observations were obtained with the 
United States Naval Observatory (USNO)
 1m telescope using BVRI filters calibrated with Landolt standards, with
the University of
Washington Manastash Ridge Observatory (MRO) 0.76m telescope with a
1024$\times$1024 pixel Ford Aerospace CCD,
using a filter
similar to the Sloan $r$ ($\lambda_{c}\sim6230$\AA) and a Harris V filter, and
with the Apache Point Observatory (APO) 3.5m using the 2048$\times$2048 pixel
SITe CCD system SPIcam with the Sloan $r$ filter. 
Follow-up spectroscopic observations
were conducted on SDSS1553,
using the Double Imaging Spectrograph (DIS) on the 
APO 3.5m telescope at low resolution ($\sim$12\AA) with a 1.5$\arcsec$ slit,
providing flux-calibrated data from 3800-10000\AA. These observations showed 
that the cyclotron features are highly variable throughout the orbital period (Figure 3).
Finally, to confirm the suspected magnetic 
nature of the objects, 
circular polarization observations were obtained with the CCD Spectropolarimeter
SPOL on the 6.5m MMT and on the Steward Observatory 2.3m telescope (Figures
4 and 5).
SPOL was used with a low-resolution grating and a 1.1$\arcsec$ slit,
providing spectral
coverage of $\sim$4200$-$8400\AA\ at a resolution of $\sim$15\AA.
Observations for both objects are summarized in Table 1.

\section{SDSS1553}

The nights of MRO and USNO photometry (Figure 2) reveal periodic,
 highly modulated light
curves. The spectrum (Figure 1) indicates that the $r$, $R$ and $V$ filter 
passbands are dominated by the 
strong 6200\AA\
harmonic, while the $B$ filter
contains the 4600\AA\ harmonic. The $I$ band
shows only a 0.15 mag modulation, which may be due to a
harmonic near 9200\AA\ (see below).
The best period
determined from combining the 5 nights of $r$ data is 0.18297$\pm$0.00004d
(4.39 hr)
and Figure 2 shows the data phased on this period (with arbitrary zero phase
using the first photometric data point at JD2452164.73522).
The sinusoidal shape of
the light curve implies we are seeing the geometrical change associated with
the changing viewing angle of a magnetic pole. While the overall sinusoidal
shape of the $V$, $R$ and $r$ light curves suggests that the pole is not self-eclipsed
by the white dwarf i.e. $i$ (angle of
rotation axis
to the line of sight) + $\beta$ (angle between rotation axis and the
magnetic pole)
$<$90$^{\circ}$, the dip in the $B$ light curve at phase 0.2 might indicate
cyclotron beaming so there could be a grazing eclipse of part of the accretion
spot at phase 0.7.

With this period, we were able to phase the spectroscopic data into an
orbital sequence (Figure 3) which shows the changing amplitude of the
cyclotron features.
The spacing, the large amplitudes, and the narrow and asymmetric profiles of 
these
features are all
indicative of cyclotron emission at low electron temperatures (WF2000).
The hump locations are consistent with cyclotron harmonics n=3 (6200\AA) and n=4
(4600\AA) in a magnetic field near 58 MG. The narrow widths imply T$_{e}<$5
keV. The circular polarization spectrum (Figure 4) confirms this conclusion by
showing that the features are highly polarized and the large difference in
polarization between the 6200\AA\ and 4600\AA\ features
shows that the emission changes from marginally 
optically thick at n=3 to thin at n=4.

It is clear that SDSS1553 has accretion characteristics that are unlike
the majority of polars (WD2000). 
We derived an upper limit of 0.04 c/s from the RASS, and this lack of strong
X-rays supports a low accretion rate, while the
lack of strong Balmer emission lines 
indicates a greatly reduced ionizing UV flux (Liebert et al. 1978). 
The weak and narrow Balmer emission may originate from the irradiated secondary, as is common in polars with low
mass transfer (Schmidt et al. 1981), but time-resolved spectropolarimetry at
higher spectral resolution is needed to determine the exact viewing
and magnetic geometry. While the cyclotron features of most polars show
maximum strength when viewed perpendicular to the field lines and maximum
circular polarization when viewed along the field lines, the angle dependence
of the two is expected to be more similar at low optical depth and lower
harmonic number (WF2000). This is borne out, in part, in the low $\dot{M}$
system RX J1007.5-2016 (Reinsch et al.
(1999) where the cyclotron features are at highest
intensity when viewed along the field
lines.

While many polars show temporary low accretion states, the cyclotron humps 
present during those low states still indicate high temperature and
optical depth (e.g. VV Pup; 
Visvanathan \& Wickramasinghe 1979). SDSS1553 appears to belong to a rare
group of polars with extremely low accretion rates 
($\sim$10$^{-13}$M$_{\odot}$ yr$^{-1}$), 
plasma temperatures ($<$5 keV), and specific accretion rates 
(10$^{-3}$ g cm$^{-2}$ s$^{-1}$) as described by Schwope et al. 
(1999).
These conditions place these polars  
within the bombardment solution of heating by particle collisions rather
than shocks. SDSS1553 and HS 0922+1333 (Reimers \& Hagen 2000) appear to be
the most extreme members of this group, based on their lack of X-rays and
similarity of cyclotron features, while the others (WX LMi, 
RX J012851.9-233931, RX J1007.5-2016) have some X-ray emission
and/or weaker and broader cyclotron harmonics.  

Using SDSS template stars of late spectral class (Hawley et al. 2002), the TiO
band strengths were used to identify an M5V star (to within one spectral class)
in SDSS1553. Using
the mean $i-J$ color of +2.73 for M5 stars in the Early Data Release 
together
with an absolute $J$ magnitude of +9.38 and a measured Sloan $i$ of +17.2,
we infer a distance modulus for SDSS1553 in 
Sloan $i$ of +5.1, or a distance of 100 pc. However, it should be noted that
several studies (Friend et al. 1990, Beuermann et al. 1998, Harrison et al.
2000) have found that the secondary stars in cataclysmic variables may not
be like ZAMS stars.

Subtracting the M5V secondary from
the APO spectra reveals the expected n=2 cyclotron harmonic near $\sim$9200\AA.
With the secondary as well as the 
 large cyclotron features removed,
the remaining flux was matched to the flux of DA white dwarfs at
various temperatures (Hubeny \& Lanz 1995), with radius 
of 8$\times 10^{8}$ cm and distance of 100 pc. This gave
an upper limit of 10,000K to the temperature
of the underlying white dwarf, assuming no continuum contribution from the
cyclotron emission. Polars with periods $>$ 3 hrs generally have white
dwarfs with temperatures $\ge$20,000K (Sion 1999).
Thus, it appears that SDSS1553 contains
a very cool white dwarf or, alternatively, its radius could be unusually
small ($M\gtrsim0.6 M_{\odot}$). 
It is interesting that HS 0922+1333, with a period of 4.1 hrs, also appears
to contain a cool white dwarf (Reimers \& Hagen 2000).

\section{SDSS1324}

Although SDSS1324 is a much fainter object, with greater noise in the spectrum
and spectropolarimetry, it appears to present the most extreme example
of a low accretion-rate system of all known so far.
The spectrum
(Figure 1) is dominated by a large amplitude, narrow cyclotron feature
near 5700\AA\ and a second possible feature at 4250\AA. These could be
the 3rd and 4th harmonics in a field near 63 MG. The spectropolarimetry
(Figure 5) shows the 5700\AA\ feature is highly circularly polarized. 
As in SDSS1553, the
$r$ light curve of SDSS1324 (Figure 6) shows a sinusoidal modulation of high
amplitude (1.3 mag amplitude) on roughly a 2.6 hr timescale, 
likely indicating a
large modulation of the cyclotron feature throughout the orbit. Once again,
the underlying contribution from the white dwarf must be very small.
 We derive an
upper limit of 0.02 c/s from the RASS, even lower than SDSS1553.

\section{Conclusions}

Our photometry, spectroscopy and polarimetry of the SDSS source
SDSS1553 have revealed a polar system with an orbital period of 4.39 hr.
The spectrum is dominated by extreme amplitude, highly polarized
 cyclotron harmonics near
6200\AA\ and 4600\AA, 
indicating a white dwarf magnetic field strength of 58 MG, and TiO
features from an M5V secondary star, indicating a distance of 100 pc.
Similar cyclotron features and photometric variability  in SDSS1324 indicate 
a polar with an
orbital period near 2.6 hr. 
The narrowness
and extreme amplitude of the cyclotron features imply that these systems are
in the regime of low plasma temperature and very low specific accretion 
rate (the bombardment solution) where the accreting area is heated by
particle collisions and the accretion luminosity appears as cyclotron
radiation. 
The low count rates in the
RASS ($<$0.04 c/s) support this view.  With its ability to probe a wide 
variety of stellar
systems, the SDSS is contributing to a less
biased view of the conditions in polars, especially at
low mass transfer rates.

\acknowledgements
Funding for the creation and distribution of the SDSS Archive has been provided by the Alfred P. Sloan Foundation, the Participating Institutions,
the National Aeronautics and Space Administration, the National Science 
Foundation, the U.S. Department of Energy, the Japanese
Monbukagakusho, and the Max Planck Society. The SDSS Web site is 
http://www.sdss.org/. Studies of magnetic stars and stellar systems at Steward
Observatory is supported by the NSF through AST 97-30792. 
The SDSS is managed by the Astrophysical Research Consortium (ARC) for the 
Participating Institutions. The Participating Institutions are The
University of Chicago, Fermilab, the Institute for Advanced Study, the 
Japan Participation Group, The Johns Hopkins University, Los Alamos
National Laboratory, the Max-Planck-Institute for Astronomy (MPIA), the 
Max-Planck-Institute for Astrophysics (MPA), New Mexico State
University, Princeton University, the United States Naval Observatory, and the 
University of Washington.

\clearpage
\begin{deluxetable}{lcclcc}
\tablenum{1}
\tablewidth{0pt}
\tablecaption{Summary of Observations}
\tablehead{
\colhead{SDSS} & \colhead{UT Date} & \colhead{Obs} & \colhead{Data} & 
\colhead{Exp (min$\times$Num)} &
\colhead{Length (hr)} } 
\startdata
1553 & 2001 Mar 23 & SDSS & ugriz & 1$\times$1 & 0.02 \nl
1553 & 2001 May 26 & SDSS & Spectrum & 87$\times$1 & 1.5 \nl
1553 & 2001 Sept 2 & APO & DIS Spectra & 10-20$\times$2 & 0.5 \nl
1553 & 2001 Sept 3 & APO & DIS Spectra & 10-15$\times$3 & 0.6 \nl
1553 & 2001 Sept 4 & APO & DIS Spectra & 10-15$\times$2 & 0.4 \nl
1553 & 2001 Sept 12 & MRO & CCD filter r & 5$\times$5  & 2.7 \nl
1553 & 2001 Sept 12 & APO & DIS Spectra & 10$\times$3 & 0.8 \nl
1553 & 2001 Sept 14 & MRO & CCD filter r & 5$\times$10 & 5.0 \nl
1553 & 2001 Sept 15 & MRO & CCD filter r & 5$\times$57 & 5.3 \nl
1553 & 2001 Sept 15 & USNO & CCD BVRI & 1-3$\times$6 & 1.8 \nl
1553 & 2001 Sept 17 & MRO & CCD filter V & 5-10$\times$30 & 3.9 \nl
1553 & 2001 Sept 18 & APO & DIS Spectra & 10$\times$9 & 1.9 \nl
1553 & 2001 Sept 19 & APO & DIS Spectra & 10$\times$13 & 2.4 \nl
1553 & 2001 Sept 20 & MRO & CCD filter r & 5$\times$5 & 4.7 \nl
1553 & 2001 Sept 21 & MRO & CCD filter r & 5$\times$9 & 0.8 \nl
1553 & 2001 Sept 21 & USNO & CCD BVRI & 1-3$\times$7 & 2.1 \nl
1553 & 2001 Sept 25 & USNO & CCD BVRI & 1-3$\times$8 & 2.1 \nl
1553 & 2002 Feb 7 & APO & DIS Spectra & 10$\times$3 & 1.8 \nl
1553 & 2002 Feb 19 & MMT & SPOL & 24$\times$1 & 0.4 \nl
1324 & 2002 May 5 & SDSS & ugriz & 1$\times$1 & 0.02 \nl
1324 & 2002 Mar 8 & SDSS & Spectrum & 85$\times$1 & 1.4 \nl
1324 & 2002 Mar 24 & APO & CCD filter r & 10$\times$14 & 3.5 \nl
1324 & 2002 Mar 30 & APO & CCD filter r & 10$\times$21 & 4.2 \nl
1324 & 2002 May 8 & SO & SPOL & 20$\times$7 & 2.2 \nl
\enddata
\end{deluxetable}

\clearpage
\figcaption{SDSS spectra of the newly discovered Polars. SDSS1553 has been
smoothed with a running 3 point boxcar and SDSS1324 with a 9 point boxcar.} 

\figcaption{Light curves of SDSS1553
folded on the orbital period of 4.39 hrs. The Sloan r covers MRO data from
Sept. 12-21 while the BVRI data are USNO data from Sept. 15, 21 and 25.}

\figcaption{APO time-resolved spectra of SDSS1553 showing the changing amplitude of the
cyclotron humps. Phasing is arbitrary, using the start time of the MRO 
photometry
(JD2452164.73522). Spectra (top to bottom) were obtained on Sept. 3, 12,
10 and 4. To facilitate comparison of data obtained with different
transparency/seeing conditions, all spectra were normalized by setting
the (mainly) M-star continuum just shortward of 7600\AA\ to the common value
of $F_{\lambda}\sim35\times10^{-17}$ erg s$^{-1}$ cm$^{-2} \AA^{-1}$. }

\figcaption{SPOL data for SDSS1553 showing the circular polarization (top) and
total flux spectrum 
(bottom). Note the very high polarization of the broad emission features at
4600\AA\ and 6200\AA\, confirming their cyclotron nature.}

\figcaption{Phase-averaged SPOL data for SDSS1324 showing the circular 
polarization (top) and total flux spectrum
(bottom). Again, the strong polarization confirms the cyclotron nature of
the broad emission feature at 5700\AA.}

\figcaption{APO light curve of SDSS1324 in a Sloan $r$ filter on 2002 March 30,
showing a modulation at a period near 2.6 hr. 
Magnitudes were obtained relative to comparison stars on each frame and have
uncertainties of 0.03 (at 20.5 mag) to 0.1 (at 22nd mag).}
\newpage
\plotone{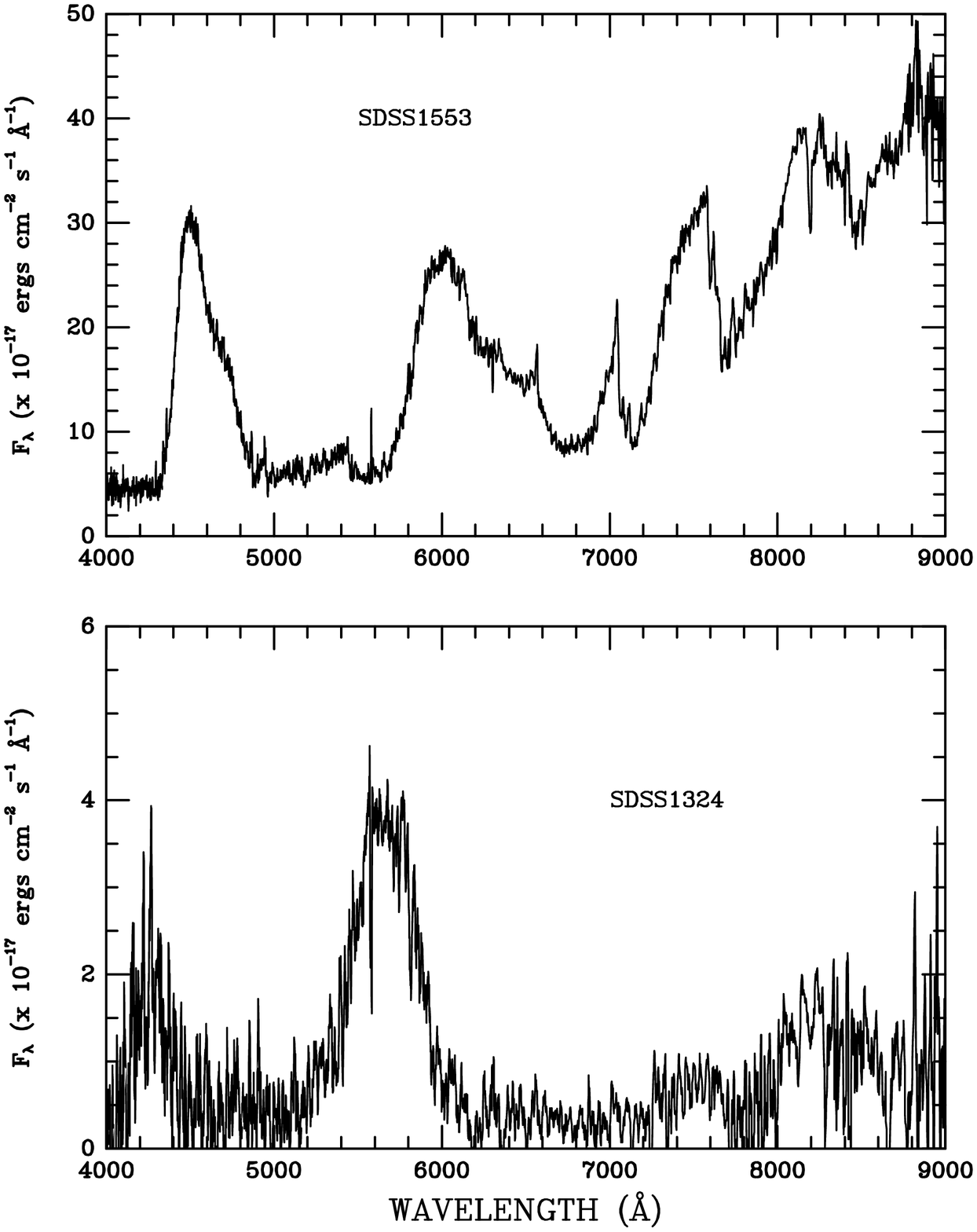}
\newpage
\plotone{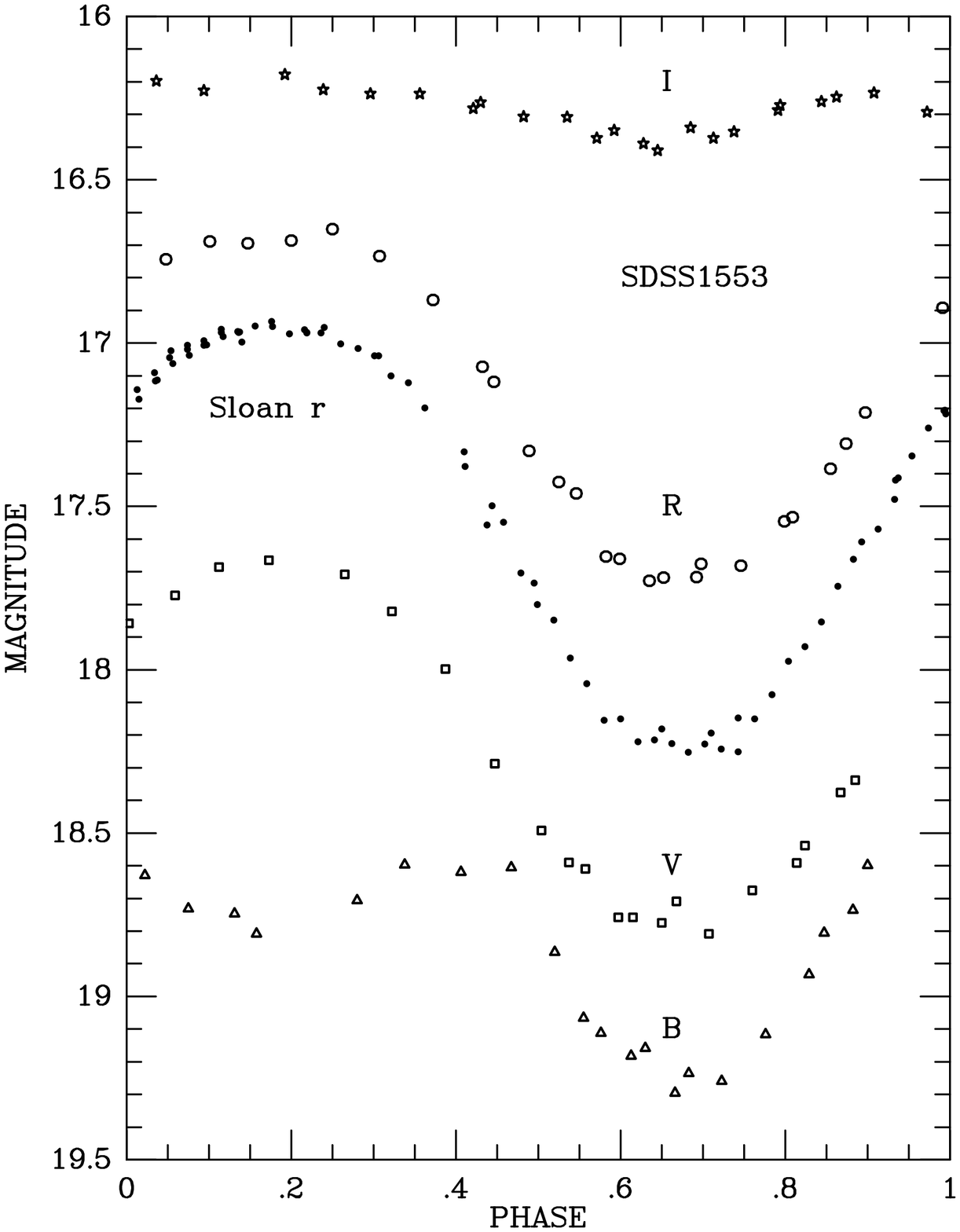}
\newpage
\plotone{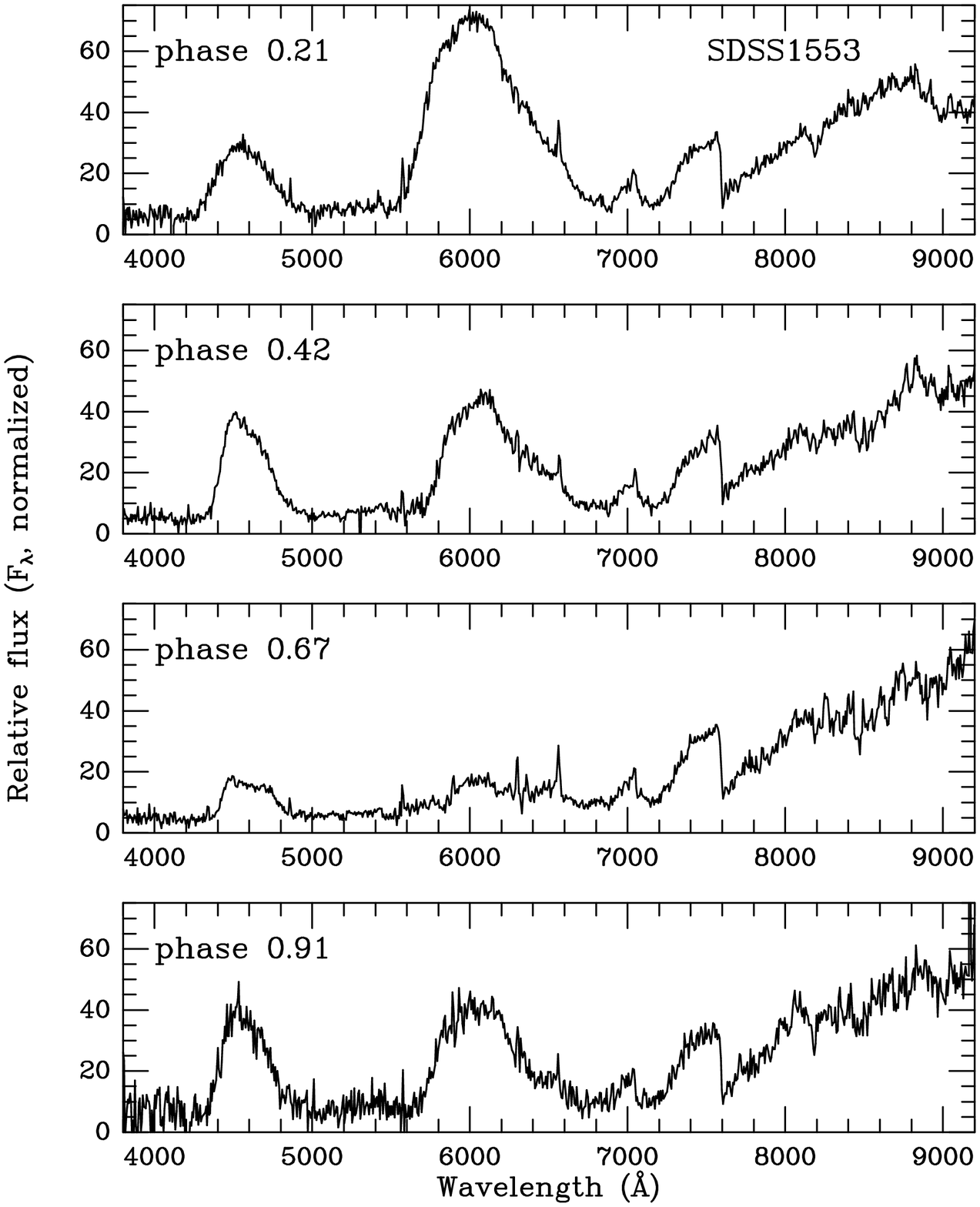}
\newpage
\plotone{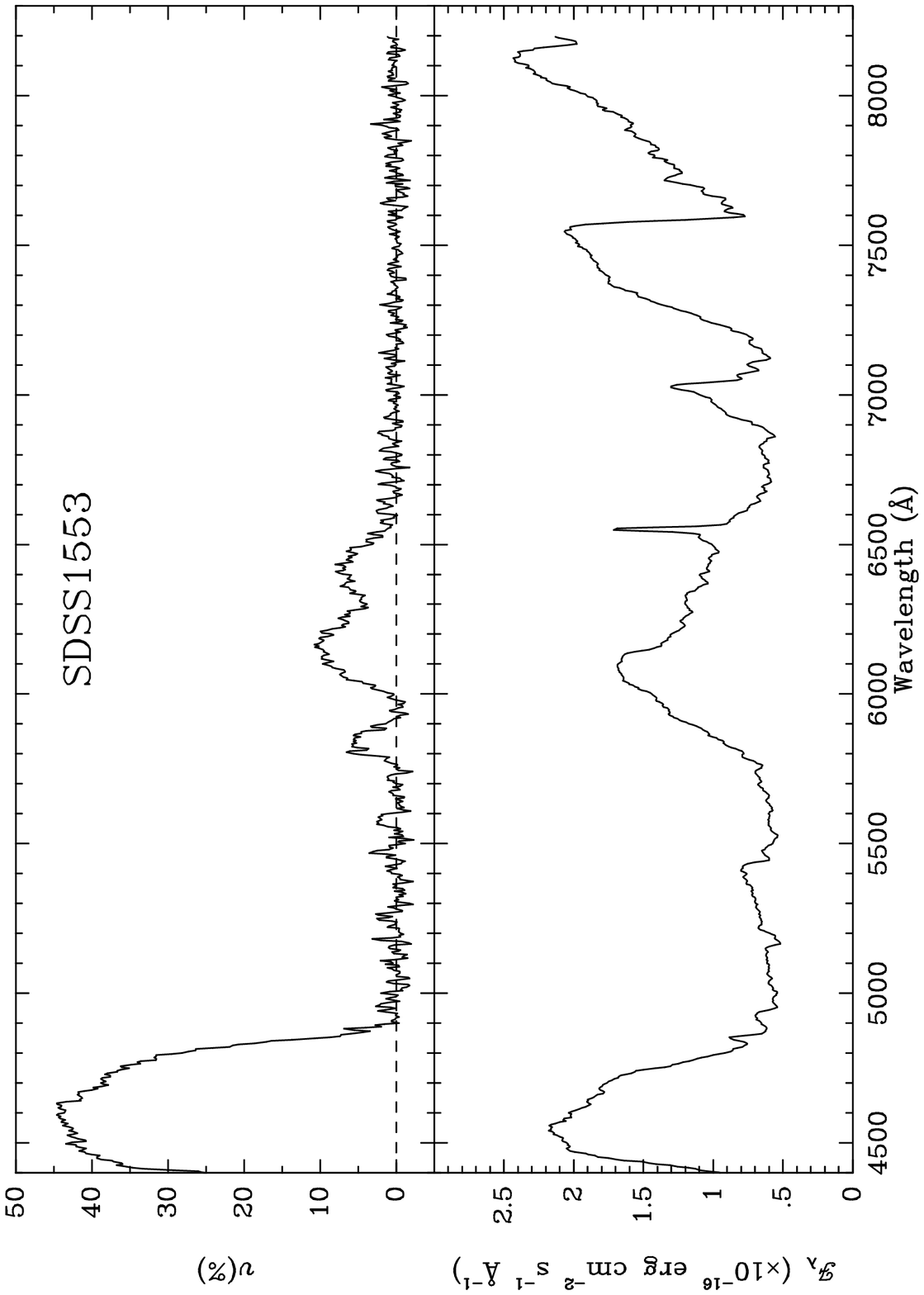}
\newpage
\plotone{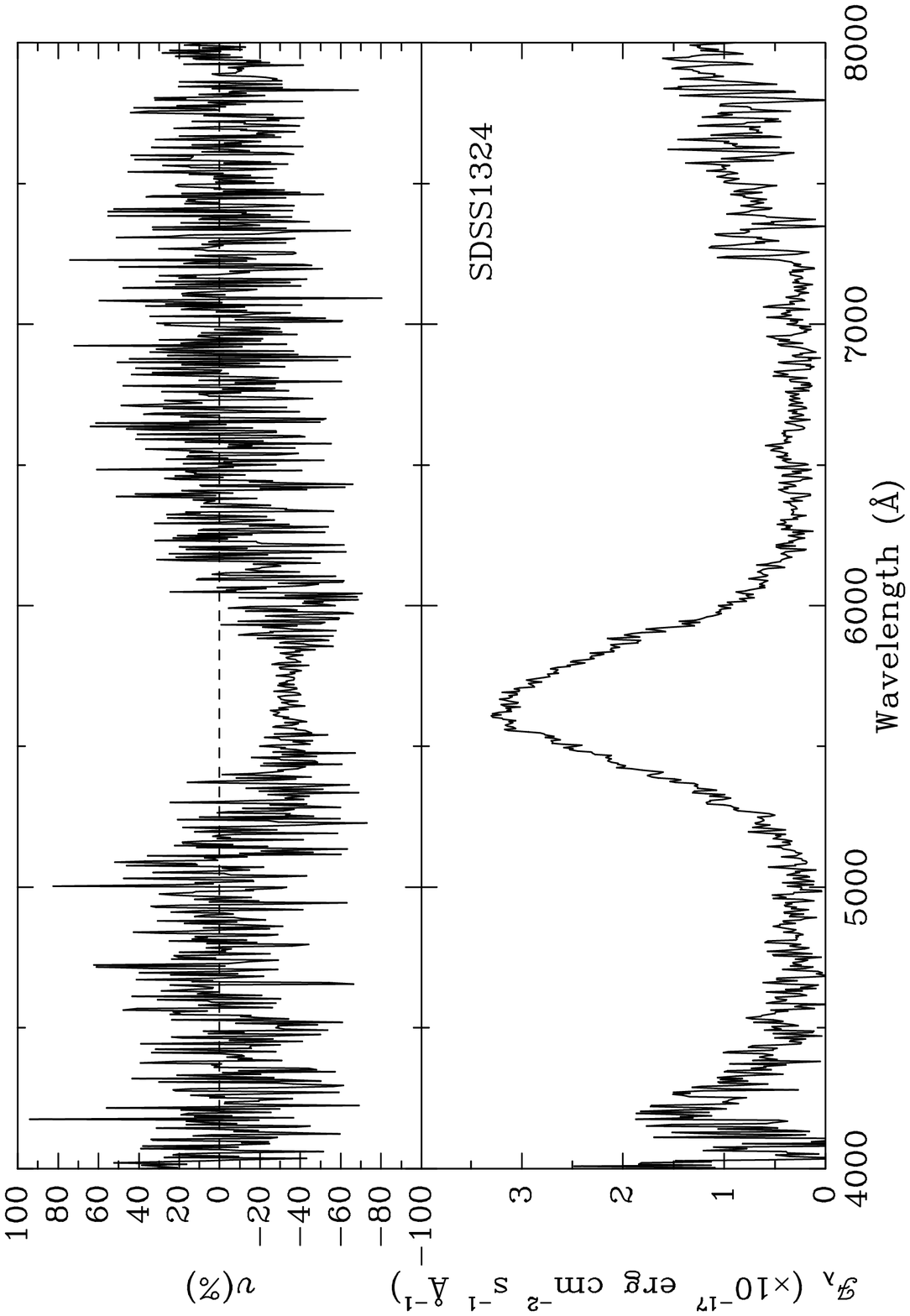}
\newpage
\plotone{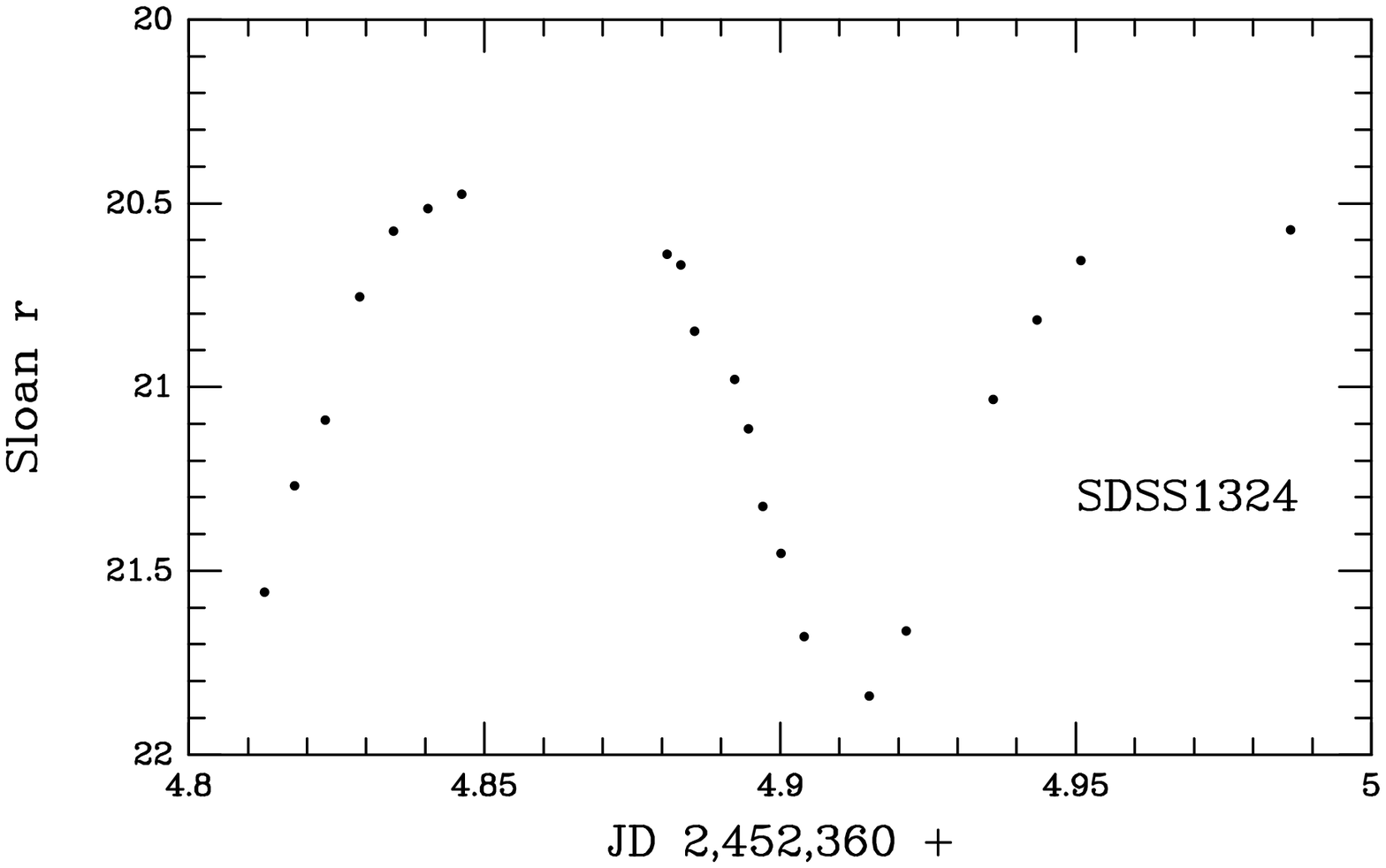}
\end{document}